\documentclass[english,journal=jacsat,manuscript=article,etalmode=truncate,maxauthors=0,layout=twocolumn,number]{achemso}
\setkeys{acs}{etalmode=truncate,maxauthors=0,articletitle=true,email=false}

\usepackage{amsmath}             
\usepackage{amssymb}             
\usepackage{wasysym}             
\usepackage{color}               
\usepackage{setspace}            
\usepackage{graphicx}            
\usepackage{svg}
\usepackage{wrapfig}             
\usepackage{array,booktabs}      

\usepackage{xr} 
\usepackage{multirow}
\usepackage{float}
\usepackage{threeparttable}
\SectionNumbersOn     

\usepackage[font=footnotesize,labelfont=bf,labelsep=period,width=0.75\textwidth]{caption}   
\usepackage[font=footnotesize,labelfont=bf,labelsep=period]{subcaption}                     
\DeclareCaptionSubType*[arabic]{figure}                                                     
\DeclareCaptionLabelFormat{subfiglabel}{Figure #2}                                          
\usepackage[capitalize]{cleveref}
\crefname{figure}{Fig.}{Figs.}
\Crefname{figure}{Figure}{Figures}
\crefname{table}{Tab.}{Tabs.}
\Crefname{table}{Table}{Tables}
\crefname{equation}{Eq.}{Eqs.}
\Crefname{equation}{Equation}{Equations}
\crefname{section}{Sec.}{Secs.}
\Crefname{section}{Section}{Sections}

\title[Running Title]{Relativistic Dirac--Coulomb--Breit  Four-Component Multireference Perturbation Theory within the Small Tensor Product Distributed Active Space Framework}
\author{Rajat Majumder}
\affiliation[University of Washington]
{Department of Chemistry, University of Washington, Seattle, WA, 98195}
\author{Shiv Upadhyay}
\affiliation[University of Washington]
{Department of Chemistry, University of Washington, Seattle, WA, 98195}
\author{Ryan A. Beck}
\affiliation[University of Washington]
{Department of Chemistry, University of Washington, Seattle, WA, 98195}
\author{Martijn Oele}
\affiliation[University of Washington]
{Department of Chemistry, University of Washington, Seattle, WA, 98195}
\author{Xiaosong Li}
\email{xsli@uw.edu}
\affiliation[University of Washington]
{Department of Chemistry, University of Washington, Seattle, WA, 98195}

\abbreviations{IR,NMR,UV}

\begin{document}

\twocolumn[
\begin{@twocolumnfalse}
\begin{abstract}
We present a four-component multireference second-order perturbation theory (4C-MRPT2) within the small tensor product distributed active space (STP-DAS) framework. The formulation is compatible with the Dirac--Coulomb (DC), Dirac--Coulomb--Gaunt (DCG), and Dirac--Coulomb--Breit (DCB) Hamiltonians, and inherits the memory-efficient and massively parallel STP-DAS algorithm, enabling perturbative treatments over very large external spaces.
Benchmark calculations on noble-gas and group-13 atoms demonstrate that 4C-MRPT2 efficiently recovers all-electron dynamic correlation while providing new insight into relativistic correlation effects. The calculations show that the Breit contribution to the correlation energy increases rapidly with atomic number and reaches approximately 4\% of the total correlation energy for Xe. The method also shows that second-order correlation energies are only weakly dependent on the choice of multiconfigurational reference orbitals, and that frozen-core and frozen-virtual approximations substantially reduce computational cost with minimal loss of accuracy.
\end{abstract}
\end{@twocolumnfalse}
]
\section{Introduction}
The predictive modeling of molecular science rests on the accurate and balanced treatment of relativistic effects and electron correlation.\cite{Pyykko88_563,Faegri07_book,Wolf15_book,Saue16_074104,Li25_4301} 
The four-component configuration interaction hierarchy provides the most rigorous treatment of relativistic electron correlation by describing spin--orbit coupling, static correlation, and dynamic correlation variationally within a unified framework. Its steep computational scaling, however, limits applications to relatively small systems.\cite{Werner11_054101,Shepard12_108} 

 
A more practical approach is to retain the relativistic multiconfigurational reference\cite{Mizukami15_4733,Li20_2975,Li20_090903,Li22_5011} while recovering dynamical correlation perturbatively. In the non-relativistic regime, second-order multireference perturbation theories, most notably complete active space perturbation theory (CASPT2) and $n$-electron valence state perturbation theory (NEVPT2),\cite{Wolinski90_5483,Roos92_1218,Koc99_2808,Hirao06_234110,Mizukami15_4733,Shiozaki16_3781} have become the methods of choice by combining a compact multiconfigurational reference with a polynomial-scaling treatment of external excitations, formally $\mathcal{O}(\mathcal{N}^{5})$ in the number of orbitals $\mathcal{N}$.\cite{Roos92_1218,Malrieu01_297,Malrieu01_10252} Extending this strategy to relativistic electronic structure is considerably more challenging because of the spinor nature of the wave function, the loss of spin symmetry, and the increased complexity of relativistic two-electron interactions. Nevertheless, substantial progress has been made with relativistic CASPT2,\cite{Kimihiko06_234110,Li22_2983,Minori25_1249} NEVPT2,\cite{Mizukami15_4733,Sokolov24_4676} algebraic diagrammatic construction (ADC),\cite{Dutta25_104106,Dutta26_3971} driven similarity renormalization group perturbation theory (DSRG-PT2),\cite{Evangelista26_7230} and generalized Van~Vleck perturbation theory (GVVPT).\cite{Liu14_1489} However, most of these developments have focused on two-component Hamiltonians, while existing four-component implementations are limited to the Dirac--Coulomb Hamiltonian. A four-component multireference perturbation theory that explicitly incorporates Breit interactions therefore remains an important outstanding challenge.
 
In this work, we develop a four-component multireference second-order perturbation theory (4C-MRPT2) that treats the Dirac--Coulomb--Breit Hamiltonian consistently at both the multiconfigurational reference and second-order perturbation levels. The method builds upon our recent advances in efficient Dirac--Coulomb--Breit integral evaluation and transformation,\cite{Li21_3388,Li22_064112,Li23_171101} together with fully variational four-component multiconfigurational electronic structure theory.\cite{Li23_044101} It is further integrated with the small tensor product distributed active space (STP-DAS) framework,\cite{Li24_041404,Li25_11016} which overcomes the memory bottleneck associated with extreme-scale configuration interaction calculations. Beyond its distributed-memory capabilities, STP-DAS provides a flexible framework for orbital-space partitioning and excitation restrictions, enabling an efficient implementation of the second-order perturbative correction.


\section{Relativistic Multireference Perturbation Theory within the STP-DAS Framework}
\label{theory:main}
In this section, the following notations are used, unless otherwise specified:
\begin{itemize}
    \item $p, q, r, s$: molecular orbitals (MOs)
\item The components of the $\boldsymbol{\alpha}$ vector are defined as
\begin{equation}
    \boldsymbol
    {\alpha}_{i,J} = 
    \begin{pmatrix}
    {\bf 0}_2 & \boldsymbol\sigma_J \\
    \boldsymbol\sigma_J & {\bf 0}_2
    \end{pmatrix},
    \quad J = \{x, y, z\}\notag
\end{equation}
with the $\boldsymbol\sigma$ vector consisting of Pauli matrices. 
\begin{align}
\mathbf{I}&=\begin{pmatrix} 1 & 0\\ 0&1 \end{pmatrix},
\boldsymbol\sigma_x=\begin{pmatrix} 0 & 1\\ 1&0 \end{pmatrix}, \notag\\
\boldsymbol\sigma_y&=\begin{pmatrix} 0 & -i\\ i&0 \end{pmatrix}, \boldsymbol\sigma_z=\begin{pmatrix} 1 & 0\\ 0&-1 \end{pmatrix}.\notag
\end{align}
\item $\mathcal{I}^+,\mathcal{J}^+,\mathcal{K}^+,\mathcal{L}^+$: four-spinor configurations in the positive-energy space
\item $\mathbb{P}^+, \mathbb{Q}^+$: configurational space partitions, consisting of four-spinor configurations
\item $h^\text{FC}_{pq} = h_{pq} + \sum_{i'}(g_{pqi'i'} - g_{pi'i'q})$
\item $h'_{pq} = h_{pq}^\text{FC} - \frac{1}{2}\sum_r g_{prrq}$
\end{itemize}

\subsection{Four-Component Relativistic Framework and No-Virtual Pair Approximation}
\label{theory:4c}

In the Coulomb gauge, the instantaneous Dirac--Coulomb--Breit two-electron operator is
\begin{align}
V_{ee} &= \sum_{i=1}^N \sum_{j>i} (g^C(i,j)+g^B(i,j))\label{eq:dcb}\\ 
    g^C(i,j) &= \frac{1}{r_{ij}} \label{eq:Coulomb}\\
    g^B (i,j) &= - \frac{\boldsymbol{\alpha}_i \cdot \boldsymbol{\alpha}_j}{r_{ij}} + \frac{(\boldsymbol{\alpha}_i \times {\bf r}_{ij}) \cdot ( \boldsymbol{\alpha}_j \times {\bf r}_{ij}) }{2r^3_{ij}} \label{eq:breit}
\end{align}
where the Breit term, \cref{eq:breit}, includes the Gaunt interaction (the first term) and an additional gauge term.

The molecular four-component spinors are defined as,
\begin{align}
\psi_p = \begin{pmatrix}
    \phi^L_p \\
    \phi^S_p
\end{pmatrix}.
\end{align}
The large- and small-component, $\phi^L_p$ and $\phi^S_p$, are:
\begin{align}
    \phi_{p}^{L} &= \sum_{\tau\in\{\alpha,\beta\}}\sum_{\mu=1}^{N_b^L} c_{\mu\tau,p}^{L} \chi_{\mu\tau}^{L} , \label{eq:Lspinor}\\
    \phi_{p}^{S} &= \sum_{\tau\in\{\alpha,\beta\}}\sum_{\mu=1}^{N_b^S} c_{\mu\tau,p}^{S} \chi_{\mu\tau}^{S} 
\end{align}
where $\phi^L_p$ and $\phi^S_p$ are the two-spinors, with $N_b$ being the number of basis functions.

The small component basis is obtained using the restricted kinetic balance (RKB) condition,\cite{Faegri07_book,Liu10_1679,Wolf15_book,Li21_207}
\begin{align}
\chi_{\mu\tau}^{S} = \frac{1}{2c}\boldsymbol{\sigma}\cdot \boldsymbol{p} \ \chi_{\mu\tau}^{L}
\end{align}
where $c$ and $\boldsymbol{p}$ are the speed of light and linear momentum operator, respectively. 
Introducing the RKB condition into the Dirac equation results in the four-component Dirac formalism in matrix representation in atomic units:\cite{Faegri07_book}
\begin{align}
     &\begin{pmatrix}
         {\bf F}^{LL} & {\bf F}^{LS} \\
         {\bf F}^{SL} & {\bf F}^{SS}
     \end{pmatrix}
     \begin{pmatrix}
         {\bf c}^+_L &{\bf c}^-_L  \\
         {\bf c}^+_S &{\bf c}^-_S
     \end{pmatrix}
     \notag\\
     &=
     \begin{pmatrix}
         \mathbf{I}\otimes{\bf S} & {\bf 0} \\
         {\bf 0} & \mathbf{I} \otimes \frac{1}{2c^2}{\bf T}
     \end{pmatrix}
     \begin{pmatrix}
         {\bf c}^+_L &{\bf c}^-_L  \\
         {\bf c}^+_S &{\bf c}^-_S
     \end{pmatrix}
     \begin{pmatrix}
         \boldsymbol{\epsilon}^+ &{\bf 0}  \\
         {\bf 0} &\boldsymbol{\epsilon}^-
     \end{pmatrix}
     \label{eq:fourcomp}
\end{align}
where $\mathbf{F}$, $\mathbf{T}$, and $\mathbf{S}$ are the Fock, kinetic energy, and overlap matrices, respectively. 

The solution of \cref{eq:fourcomp} consists of sets of positive/negative eigenvalues ($ \{ \epsilon^+ \}$, $ \{ \epsilon^- \}$) with corresponding molecular orbital coefficients $\begin{pmatrix} {\bf c}^+_{L} \\ {\bf c}^+_{S} \end{pmatrix}$ and $\begin{pmatrix} {\bf c}^-_{L} \\ {\bf c}^-_{S} \end{pmatrix}$ for the positive/negative energy solutions.
These orbitals represent mean-field, spin-coupled, relativistically corrected molecular four-spinors and serve as the one-electron basis for subsequent correlated calculations.

In the present work, we adopt the conventional no-virtual-pair approximation, in which only the positive-energy molecular spinors,
$\begin{pmatrix}
{\bf c}^{+}_{L}\\
{\bf c}^{+}_{S}
\end{pmatrix}$,
are included in the multiconfigurational expansion and perturbation treatment, while the negative-energy orbitals are excluded from the correlated treatment.\cite{Sucher80_348,Faegri07_book,Saue16_074104} The effects of relaxing this constraint is later discussed in \cref{res:0}, where positive-negative rotations are facilitated in the variational multiconfigurational treatment.

\subsection{Four-Spinor Orbital Optimizations within No-Virtual-Pair Approximation}
\label{theory:nvpa}
Within the no-virtual-pair approximation, the correlated wave function is constructed exclusively from positive-energy molecular spinors. However, because the kinetic-balance condition couples the large- and small-component spaces, the positive- and negative-energy solutions are not strictly independent. This raises a nontrivial question regarding the proper treatment of orbital relaxation in relativistic multiconfigurational methods.

In our previous work,\cite{Li23_044101} we considered three levels of orbital optimization: (i) 4C-CASCI, in which the orbitals are kept fixed; (ii) 4C-CASSCF$^{+}$, where orbital rotations are restricted to the positive-energy spinor manifold; and (iii) 4C-CASSCF$^{\pm}$, where orbital rotations are permitted over the entire four-component spinor space. Note that in 4C-CASSCF$^{\pm}$, the negative-energy spinors are not included in the configuration expansion but are allowed to participate in orbital optimization, acting as an auxiliary virtual space for variational relaxation of the positive-energy orbitals.

We refer readers to Ref. \citenum{Li23_044101} for detailed implementation and benchmarking of the 4C-CASSCF methodology, including the treatment of orbital optimization, spinor rotations, and active-space construction within the four-component framework.


\subsection{Multireference Perturbation Theory}
\label{theory:mrpt2}
Although the underlying four-spinor orbitals may be optimized using different rotation schemes, as discussed in the previous section, the no-virtual-pair approximation restricts the many-electron Hilbert space to determinants constructed exclusively from positive-energy spinors. For a system with $N_o^+$ positive-energy orbitals and $N_e^+$ electrons, the complete configuration space contains
$\binom{N_o^+}{N_e^+}$
Kramers-unrestricted Slater determinants, $\{\mathcal{I}^+,\mathcal{J}^+,\mathcal{K}^+,\mathcal{L}^+\}$, corresponding to all possible occupations of $N_e^+$ electrons among the $N_o^+$ positive-energy spinors. 

Within the Epstein--Nesbet partitioning framework, the configurational space is divided into a primary space, $\textbf{P}^+$, and an external space, $\textbf{Q}^+$. The primary space spans the zeroth-order reference states,
\begin{align}
|\Psi_m^{(0)}\rangle &= \sum_{\mathcal{I}^+\in\mathbf{P}^+}
C^{(0)}_{m,\mathcal{I}^+} |\mathcal{I}^+\rangle, \\
\langle\Psi_m^{(0)}|\Psi_n^{(0)}\rangle &= \delta_{mn}.
\end{align}
The external space, $\textbf{Q}^+$ contains all configurations outside $\textbf{P}^+$ that couple to the reference states through the Hamiltonian. This space constitutes the first-order interacting space and is responsible for the second-order energy correction. In the present work, $\textbf{Q}^+$ is composed of all single- and double-excitation configurations that are not included in the primary space, \emph{i.e.}, 4C-MRPT2.

Based on this model partitioning, the zeroth-order Hamiltonian ($\hat{H}^{(0)}$) can be defined as,
\begin{align}
\label{eq:zeroh}
\hat{H}^{(0)} &= \sum_{\cal{I}^+,\cal{J}^+ \in \mathbf{P}^+} |{\cal{I}^+}\rangle \langle{\cal{I}^+}|\hat{H}|{\cal{J}^+}\rangle \langle{\cal{J}^+}| \nonumber  \\ &+  \sum_{\cal{K}^+ \in \mathbf{Q}^+} |{\cal{K}^+}\rangle \langle{\cal{K}^+}|\hat{H}|{\cal{K}^+}\rangle \langle{\cal{K}^+}|
\end{align}
where $\hat{H}^{(0)}$ is block-diagonal within reference $\textbf{P}^+$ space and diagonal within the external $\textbf{Q}^+$ space. Using this model-space partitioning, we arrive at the expression for obtaining second-order energy correction, $E^{(2)}_m$ to reference state ${|\Psi^{(0)}_m}\rangle$,
\begin{align}
\label{eq:pt2}
E^{(2)}_m &= \sum_{\cal{K}^+ \in \mathbf{Q}^+} \frac{\langle{\mathcal{K}^+}|\hat{H}|{\Psi^{(0)}_m}\rangle \langle{\Psi^{(0)}_m}|\hat{H}|{\cal{K}^+}\rangle} {E^{(0)}_m - E_{\cal{K}^+\cal{K}^+}}
\end{align}
where $ E_{\cal{K}^+\cal{K}^+} = \langle{\cal{K}^+}|\hat{H}|{\cal{K}^+}\rangle $ and $E^{(0)}_m$ is the reference energy of state ${|\Psi^{(0)}_m}\rangle$. 

While the working equations of 4C-MRPT2 closely resemble those of their non-relativistic one-component and relativistic two-component counterparts, the underlying molecular orbital integrals are fundamentally different, as they are obtained from atomic orbital integrals over the Dirac--Coulomb--Breit Hamiltonian.

\subsection{STP-DAS Enabled 4C-MRPT2}

While extending a relativistic multireference method to include a one-shot second-order perturbative correction is conceptually straightforward, the primary challenge lies in the efficient traversal of the enormous determinant space associated with external excitations. This challenge becomes particularly acute for large orbital spaces, where the number of external determinants can exceed the size of the reference space by many orders of magnitude. In this work, we develop an STP-DAS-enabled MRPT framework that enables scalable multireference perturbation theory calculations for previously inaccessible problem sizes.

The central idea of STP-DAS is to decompose the CI problem, specifically the $\boldsymbol{\sigma}$-build, into a hierarchy of small tensor products, each possessing a unique global address determined analytically from the distributed active-space configuration. This representation provides a structured organization of the CI space while enabling efficient storage and manipulation of extremely large determinant expansions. We extend this framework towards enabling STP-DAS multireference perturbation theory which we elaborate below.

\begin{figure}[t]
    \centering
    \captionsetup{width=0.5\textwidth}
    \includegraphics[width=0.5\textwidth]{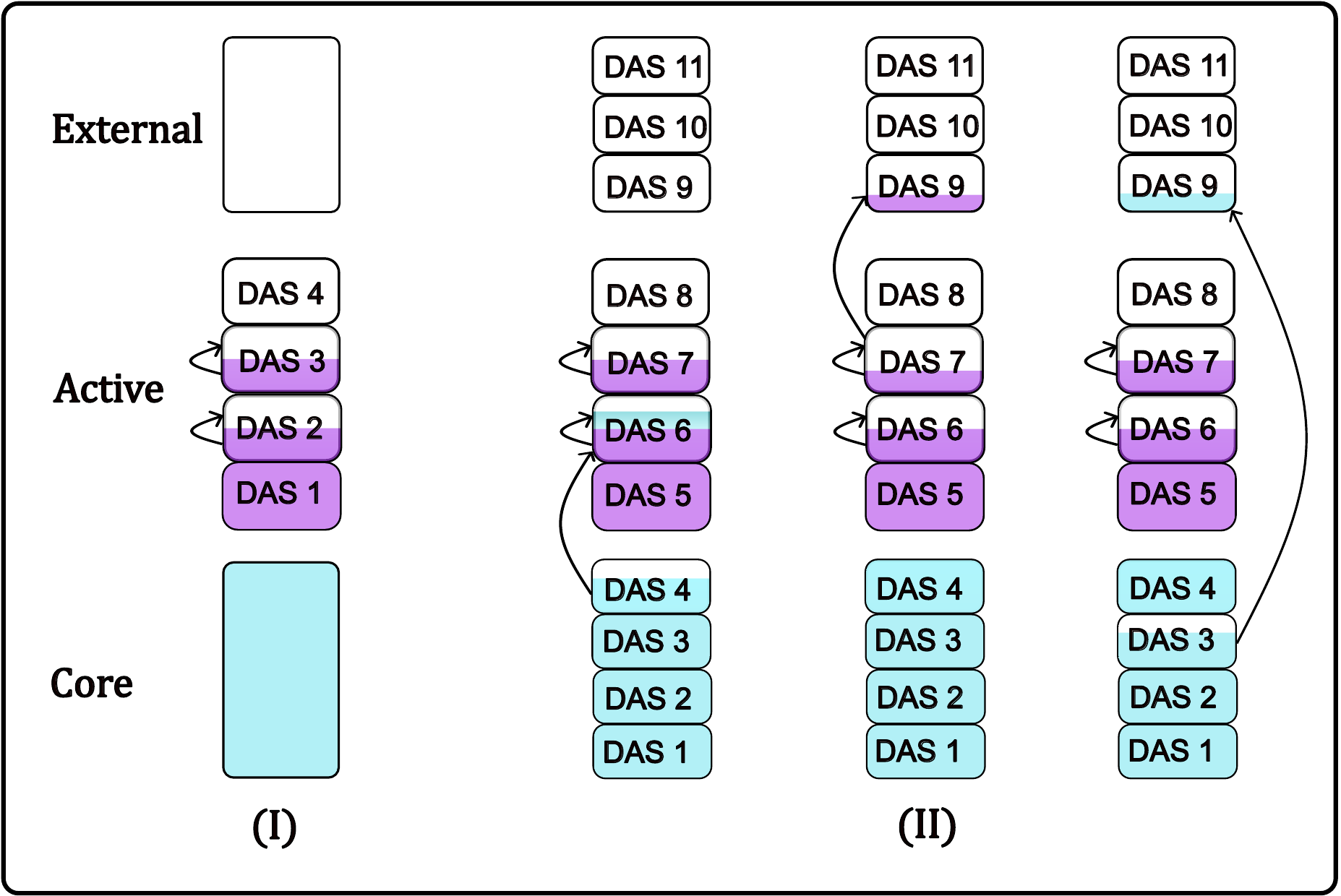}
    \caption{Orbital partitioning in relativistic MRPT2 within the STP-DAS framework where the boxes symbolize DAS spaces and the correlated regions involved in electronic excitations. (I) Elucidates the formation of configuration categories in CI where electronic excitations are limited to the active region. (II) Elucidates the formation of configuration categories belonging to the external ($\mathbf{Q}^+$) space in MRPT2, where single and double excitations connecting the external and reference spaces are taken into account. Note that the DAS partitioning within the correlated space encompasses positive energy spinors that are not frozen, invoking no virtual-pair approximation (NVPA).}
    \label{fig:pt2_das}
\end{figure}

Configurations in the primary reference space are denoted as $\mathbf{P}^+$, arising from electronic excitations only within the active space. Configurations in the external space, $\mathbf{Q}^+$, arise from excitations of electrons from the core space to the active space, from the active space to the external virtual space, and from the core space directly to the external virtual space. To facilitate the small-tensor product (STP) algorithm, the core and external virtual orbital spaces are further partitioned into additional DAS subspaces, as illustrated in \cref{fig:pt2_das}, denoted by $\mathbb{Q}^+_\mu$, $\mathbb{Q}^+_\nu$, and so forth.

Inter-DAS excitations give rise to distinct configuration categories that serve as the fundamental organizational units of the STP-DAS framework. Each category, $\mathcal{A}=\{\mathbb{Q}_\mu^\mathcal{+A}, \mathbb{Q}_\nu^\mathcal{+A}, \ldots\}$, is uniquely characterized by its electron occupation pattern across the DAS subspaces, while the orbital partitioning itself remains fixed. Each DAS within a configuration category represents a local complete active space and is composed of a set of sub-determinants,
$\mathbb{Q}_\mu^{+\mathcal{A}}
=
\left\{
\mathbb{Q}_{\mu,{1}}^{+\mathcal{A}},
\mathbb{Q}_{\mu,{2}}^{+\mathcal{A}},
\mathbb{Q}_{\mu,{3}}^{+\mathcal{A}},
\ldots
\right\},$
where the indices $\mathbb{Q}_{\mu,{1}}^{+\mathcal{A}}$, $\mathbb{Q}_{\mu,{2}}^{+\mathcal{A}}$, and $\mathbb{Q}_{\mu,{3}}^{+\mathcal{A}}$ label individual sub-determinants within the DAS. \Cref{fig:pt2_das} presents representative examples of the three types of configuration categories that arise in the PT2 formalism. This categorization enables a compact tensor-product representation and efficient traversal of the external configuration space.

A full-space determinant in category $\mathcal{A}$ can be constructed using sub-determinants from each DAS space
\begin{equation}
|{\mathcal{K}^\mathcal{+A}}\rangle=|{\mathbb{K}^\mathcal{+A}_{\mu,\mathcal{K}}}\rangle\oplus|{\mathbb{K^\mathcal{+A}_{\nu,\mathcal{K}}}\rangle}\oplus\cdot\cdot\cdot    \label{eq:DASK}
\end{equation}
where $\mathbb{K}^{+\mathcal{A}}_\mu$ denotes sub-determinants in DAS space, $\mu$.

This small-tensor representation of a configuration allows the MRPT2 calculation to be reformulated as a sequence of tensor loops, where the $\boldsymbol\sigma$-build can be reinforced to obtain the interaction between the external determinants and the reference determinants for a CI/CAS electronic state, $m$. 
\begin{align}
\label{eq:sumsigma}
\sigma_{m, \cal{K}^{+\mathcal{A}}} &= {}^{1e}\sigma_{m,\cal{K}^{+\mathcal{A}}} + {}^{2e}\sigma_{m,\cal{K}^{+\mathcal{A}}} \\
{}^{1e}\sigma_{m,\cal{K}^{+\mathcal{A}}} &= \sum_{{}^+\mathcal{C} \in \boldsymbol{P}^+}
  \sum_{\mathbb{L}^{+\mathcal{C}}_{\mu}\oplus\mathbb{L}^{+\mathcal{C}}_{\nu}}
  \sum_{pq}
  P_{\mu\nu}\,
  \delta_{\bar{\mathbb{X}}^{+\mathcal{A}}_{\mu\nu}\bar{\mathbb{X}}^{+\mathcal{C}}_{\mu\nu}}\, 
  h'_{pq}\, \nonumber \\ &
  \bigl\langle \mathbb{K}^{+\mathcal{A}}_{\mu}\oplus\mathbb{K}^{+\mathcal{A}}_{\nu}
  \bigr| \hat{E}_{pq} \bigl|
  \mathbb{L}^{+\mathcal{C}}_{\mu}\oplus\mathbb{L}^{+\mathcal{C}}_{\nu} \bigr\rangle\,
  C^{(0)}_{m,\cal{L}^{+\mathcal{C}}} \\
{}^{2e}\sigma_{m,\cal{K}^{+\mathcal{A}}} &= \frac{1}{2}
  \sum_{{}^+\mathcal{B}{}^+\mathcal{C}}
  \sum_{\mathbb{J}^{+\mathcal{B}}_{\mu}\oplus\mathbb{J}^{+\mathcal{B}}_{\nu}}
  \sum_{\mathbb{J}^{+\mathcal{B}}_{\kappa}\oplus\mathbb{J}^{+\mathcal{B}}_{\lambda}}
  \sum_{\mathbb{L}^{+\mathcal{C}}_{\kappa}\oplus\mathbb{L}^{+\mathcal{C}}_{\lambda}}
  \sum_{pqrs} \nonumber \\ & P_{\mu\nu} P_{\kappa\lambda}\delta_{\bar{\mathbb{X}}^{+\mathcal{A}}_{\mu\nu}\bar{\mathbb{X}}^{+\mathcal{B}}_{\mu\nu}}\,
  \delta_{\bar{\mathbb{X}}^{+\mathcal{B}}_{\kappa\lambda}\bar{\mathbb{X}}^{+\mathcal{C}}_{\kappa\lambda}}\,
  g_{pqrs} \nonumber \\ & \bigl\langle \mathbb{K}^{+\mathcal{A}}_{\mu}\oplus\mathbb{K}^{+\mathcal{A}}_{\nu}
  \bigr| \hat{E}_{pq} \bigl|
  \mathbb{J}^{+\mathcal{B}}_{\mu}\oplus\mathbb{J}^{+\mathcal{B}}_{\nu} \bigr\rangle \nonumber \\ 
  & \bigl\langle \mathbb{J}^{+\mathcal{B}}_{\kappa}\oplus\mathbb{J}^{+\mathcal{B}}_{\lambda}
  \bigr| \hat{E}_{rs} \bigl|
  \mathbb{L}^{+\mathcal{C}}_{\kappa}\oplus\mathbb{L}^{+\mathcal{C}}_{\lambda} \bigr\rangle\,
  C^{(0)}_{m,\cal{L}^{+\mathcal{C}}}
\end{align}
Here, we introduce a Kronecker $\delta$ function $\delta_{\bar{\mathbb{X}}_\mu^{+\mathcal{A}}\bar{\mathbb{X}}_\nu^{+\mathcal{B}}}$
where $\bar{\mathbb{X}}_\mu^{+\mathcal{A}}$ refers to all but $\mathbb{X}_\mu$ DASs in category $\mathcal{A}$. In other words, the $\delta$ function is non-zero only when non-excitation DASs between categories are identical. The CI expansion vectors for state $m$ are $C^{(0)}_{m,\cal{L}^{+\cal{C}}}$ where $\cal{C}$ is a configuration category belonging to the reference space, $\textbf{P}^+$. We also highlight the following definitions: $p\in \mathbb{X}^{+\mathcal{A}}_\mu$, $q\in \mathbb{X}^{+\mathcal{B}}_\nu$, $r\in \mathbb{X}^{+\mathcal{B}}_\kappa$ and $s\in \mathbb{X}^{+\mathcal{C}}_\lambda$. $P_{\mu\nu}$ is a global address offset associated with the $\mathbb{X}_\mu$ and $\mathbb{X}_\nu$ DAS spaces (see Ref. \citenum{Li24_041404}).

Using the equations above, we can formulate the MRPT2 equations using \cref{eq:sumsigma} as, 
\begin{align}
\label{eq:pt2en}
E^{(2)}_{m} = \sum_{\cal{K}^{+\mathcal{A}} \in\, \boldsymbol{Q}^{+}}
  \frac{\bigl|\sigma_{m,\cal{K}^{+\mathcal{A}}}\bigr|^{2}}
       {E^{(0)}_{m} - H_{\cal{K}^{+\mathcal{A}}\cal{K}^{+\mathcal{A}}}}
\end{align}
where the summation is over all the configuration categories, $\cal{K}^{+\cal{A}}$ belonging to external space, $\textbf{Q}^+$.

\subsection{Frozen Core and Virtual Approximations}
\label{theory:fcv}
The Hilbert space encapsulating single and double excitations outside of the reference can often be huge, leading to substantial increase in computational complexity. Also, the Hilbert space involves all the molecular spinors within the positive energy region. This can lead to memory and time bottlenecks when doing the expensive relativistic two-electron transformation from the AO basis to the MO basis. We show this in our subsequent results sections as well. Therefore, to alleviate this problem, we have invoked frozen core and virtual spinors which are excluded from the space of single and double excitations. This leads to drastic reduction in computational complexity in obtaining approximate correlation energies due to perturbative corrections. We demonstrate later that freezing energetically high-lying virtual spinors leads to negligible deviations in correlation energy. Therefore, the approximation provides a low-cost framework in obtaining correlation energies.

\subsection{Intruder-State Resolution}
MRPT2 treatment can often give rise to intruder states, where the perturbative corrections to the electronic structure are erroneous. The common reason behind this is the vanishing denominator where the determinant, $\mathcal{K}^{+\cal{A}}$ belonging to external space, $\textbf{Q}^+$ becomes degenerate with respect to reference energy, $E^{(0)}_m$ of electronic state $m$. This problem is highly dependent on the type of zeroth-order Hamiltonian employed in MRPT2 and can arise even in robust methods such as $n$-electron valence state perturbation theory (NEVPT2).\cite{Chan09_194107,Yanai24_194105}

To address the problem, we have invoked the imaginary level-shift\cite{Malmqvist97_1196,Witek12_4053,Lindh22_4814} technique where a small imaginary shift is introduced in the denominator of \cref{eq:pt2},
\begin{align}
& \left(E^{(0)}_m - E_{\mathcal{K}^{+\cal{A}} \mathcal{K}^{+\cal{A}}}\right)^{-1} \\
& \rightarrow \left(E^{(0)}_m - E_{\mathcal{K}^{+\cal{A}} \mathcal{K}^{+\cal{A}}} + i\epsilon_{m\mathcal{K}^{+\cal{A}}}\right)^{-1} \nonumber
\end{align}
where the level-shift parameter, $\epsilon_{m\mathcal{K}^{+\cal{A}}}$ is
\begin{align}
\epsilon_{m\mathcal{K}^{+\cal{A}}} = \frac{b}{E^{(0)}_m - E_{\mathcal{K}^{+\cal{A}}\mathcal{K}^{+\cal{A}}}}
\end{align}
with $b$ being a user-defined constant, usually between 0.01 and 0.1. 

\section{Results and Discussion}
\label{results:main}

All calculations were performed using a developmental version of the Chronus Quantum software package.\cite{Li20_e1436} The speed of light was set to 137.035999074 a.u., and the standard Gaussian nuclear model was employed throughout.\cite{Dyall97_207,Li21_207}

Relativistic calculations were carried out within the Kramers-unrestricted four-component and two-component framework, in which all spinor orbitals are treated as singly occupied. The transformation of two-electron integrals from the atomic-orbital basis to the molecular-orbital basis was performed using the Pauli-quaternion relativistic integral transformation algorithm.\cite{Li26_arxiv} We did not use any level-shifts for intruder-state resolution since we did not confront any divergences for the specific set of calculations.

The calculations were performed on the University of Washington's Hyak high-performance computing system using AMD EPYC 9354P processors.

\subsection{Computational Cost of 4C-MRPT2}
\label{subsec:3}

The computational cost of 4C-MRPT2 is dominated by three steps: (i) the four-component AO-to-MO integral transformation,\cite{Li26_arxiv} (ii) the construction of the $\boldsymbol{\sigma}$-vector, which evaluates the Hamiltonian couplings between the reference ($\mathbf{P}^+$) and external ($\mathbf{Q}^+$) spaces in \cref{eq:sumsigma}, and (iii) the solution of the first-order MRPT2 amplitudes, which generate the perturbative correction to the reference wavefunction as presented in \cref{eq:pt2en}.

\begin{table}[ht]
\footnotesize
\centering
\begin{tabular}{lccc}
\hline
\hline
 & \textbf{DC (s) } & \textbf{DCG (s) } & \textbf{DCB (s) } \\
\hline
\hline
T$_\text{int}$ & 3,375 &  8,903 & 14,225 \\
T$_{\boldsymbol{\sigma}}$      &   33 &    33 &    33 \\
T$_\text{PT2}$        & 1,514 &  1,513 &  1,514 \\
\hline
\noalign{\vspace{4pt}}
Total                   & 4,978 & 10,505 & 15,828 \\
\noalign{\vspace{4pt}}
\hline
\hline
\end{tabular}
\captionsetup{width=\linewidth}
\caption{CPU wall-times (in seconds) obtained for the three stages of 4C-MRPT2 using 1 compute node. T$_\text{int}$, T$_{\boldsymbol{\sigma}}$, and T$_\text{PT2}$ are the times for the AO to MO integral transformation, the $\boldsymbol{\sigma}$-build, and the PT2 amplitude steps.}
\label{tab:mpi}
\end{table}

In \Cref{tab:mpi}, the timings for the three principal computational steps are reported. The benchmark calculation was performed for the HBr molecule using the Dyall-v2z basis set. The reference ground state was obtained from a 4C-CASCI calculation with an active space of 8 electrons in 12 four-spinors, comprising 495 reference determinants. The corresponding external space ($\mathbf{Q}^+$, 184 positive-energy spinors) contains approximately $2.17\times10^{10}$ determinants.

\Cref{tab:mpi} shows that the computational costs of the $\boldsymbol{\sigma}$-vector construction and PT2 amplitude evaluation remain essentially constant as long as the correlation space is unchanged, since both steps are independent of the underlying Hamiltonian. However, the cost of the AO-to-MO integral transformation increases substantially from the DC to the DCB Hamiltonian, even though the orbital space is unchanged. This increase reflects the significantly higher cost of computing the relativistic two-electron integrals for the more complete Hamiltonians. Consequently, the integral transformation accounts for 68\% of the total 4C-MRPT2 wall time for the DC Hamiltonian, but rises to 90\% for the DCB Hamiltonian.

\subsection{Positive--Negative-Energy Orbital Rotations in Recovering Correlation}
\label{res:0}

In this section, we demonstrate the effects of negative energy spinor rotations in 4C-CASSCF on 4C-MRPT2. 4C-CASSCF energy is stationary, but not minimal, with respect to the full set of orbital rotations: it is a minimum with respect to occupied--virtual rotations within the positive-energy branch and a maximum with respect to rotations that mix positive- and negative-energy solutions.\cite{Tatewaki07_174105,Tatewaki10_124105,Lindgren13_014108,Li23_044101} To quantify the error this introduces, we have carried out two sets of calculations for each system and Hamiltonian: orbital rotations are restricted to the positive-energy orbital space, denoted 4C-CASSCF$^{+}$, and one in which positive--negative-energy rotations are allowed, denoted 4C-CASSCF$^{\pm}$. In both cases, the perturbative 4C-MRPT2 itself employs no-virtual-pair approximation (NVPA), where the external space is restricted to the positive-energy branch of the respective reference. So the difference in 4C-MRPT2 only stems from differences in reference wavefunction definition. 

\begin{figure}[t]
    \centering
    \captionsetup{width=\linewidth}
    \includegraphics[width=\linewidth]{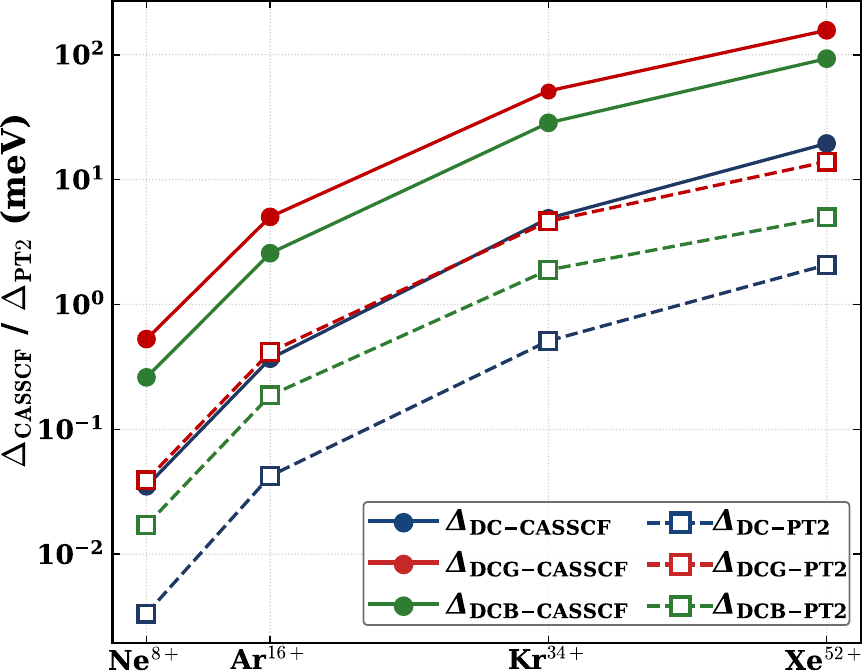}
    \caption{Effect of positive--negative-energy spinor rotations in 4C-CASSCF and 4C-MRPT2 for ionized noble gas atoms, isoelectronic with helium. The figure presents the differential 4C-CASSCF and 4C-PT2 energies (shown in \cref{eq:deltacas} and \cref{eq:deltapt}) stemming from 4C-CASSCF$^{\pm}$ and 4C-CASSCF$^{+}$ references.}
    \label{fig:negativerot}
\end{figure}

To facilitate thorough analysis, we define the following terms,
\begin{align}
\label{eq:deltacas}
&\Delta_{\text{CASSCF}} = E_{\text{CASSCF}^{\pm}} - E_{\text{CASSCF}^{+}} \\
&\Delta_{\text{PT2}} = E_{\text{PT2}^{+}}(\text{CASSCF}^\pm) 
-  E_{\text{PT2}^{+}}(\text{CASSCF}^+)\label{eq:deltapt} 
\end{align}
where $E_{\text{PT2}^{+}}(\text{CASSCF}^\pm)$ and $E_{\text{PT2}^{+}}(\text{CASSCF}^+)$ correspond to PT2 corrections within NVPA (PT2$^+$) using 4C-CASSCF$^{+}$ and 4C-CASSCF$^{\pm}$ reference orbitals, respectively. The $\Delta_{\text{CASSCF}}$ and $\Delta_{\text{PT2}}$ from \cref{eq:deltacas} and \cref{eq:deltapt} are plotted in \cref{fig:negativerot} for the isoelectronic series Ne$^{8+}$, Ar$^{16+}$, Kr$^{34+}$, and Xe$^{52+}$ at the Dirac--Coulomb (DC), Dirac--Coulomb--Gaunt (DCG), and Dirac--Coulomb--Breit (DCB) levels. An active space of two electrons and ten spinors (2e,10o), which includes the 1$s$, 2$s$ and 2$p$ spinor orbitals, was used for all atoms studied in this section with he Dyall-cv3z basis.\cite{Dyall02_335,Dyall06_441,Dyall12_1217,Dyall16_128,Dyall23_zenodo} The ${}^1\text{S}_0$ ground electronic state was assessed for this study.

Both $\Delta_{\text{CASSCF}}$ and $\Delta_{\text{PT2}}$ are positive for every system and Hamiltonian considered, indicating that freezing the positive--negative-energy orbital rotations leads to overcorrelation in four-component post-SCF calculations. As shown in \Cref{fig:negativerot}, the three Hamiltonians exhibit nearly parallel trends: although the treatment of the two-electron interaction changes the magnitude of the error by more than an order of magnitude, its dependence on nuclear charge remains essentially unchanged.

The magnitude of $\Delta_{\text{CASSCF}}$ and $\Delta_{\text{PT2}}$ increases strongly with the form of the two-electron interaction and with the atomic number. For the DC Hamiltonian, $\Delta_{\text{CASSCF}}$ increases from 0.035~meV for Ne$^{8+}$ to 19.48~meV for Xe$^{52+}$, whereas the corresponding DCB values increase from 0.26 to 96.12~meV. This behavior reflects the different operator structures of the Coulomb and Breit interactions. The Coulomb interaction is even in the Dirac matrices and does not directly couple the positive- and negative-energy states. In contrast, the Breit interaction directly couples the positive- and negative-energy states, making the Breit Hamiltonian substantially more sensitive to positive--negative-energy orbital rotations. The Gaunt Hamiltonian, often regarded as an approximate form of the Breit interaction, exhibits even larger values of $\Delta_{\text{CASSCF}}$ and $\Delta_{\text{MRPT2}}$. This behavior arises because the full Breit operator contains both the Gaunt and gauge terms, with the part of the gauge contribution canceling  one-half of the Gaunt interaction. As a result, the full Breit Hamiltonian is slightly less sensitive to positive--negative-energy orbital rotations than the Gaunt Hamiltonian alone.
 
Turning to the effect of dynamic correlation, we emphasize that the perturbative correction is formulated within the NVPA in both cases. Consequently, $E_{\text{PT2}^+}$ and $\Delta_{\text{PT2}}$ do not include any dynamic correlation arising from the negative-energy states. Such contributions lie beyond the NVPA and can only be captured through a quantum electrodynamics (QED) treatment.\cite{Liu23_e1652} $\Delta_{\text{PT2}}$ instead measures the dependence of the second-order correlation energy on the choice of reference orbitals.

\Cref{fig:negativerot} shows that the PT2 correlation energy within the NVPA depends only weakly on how the reference orbitals are optimized. Despite the substantially larger correlation space, the difference in the PT2 correlation energy obtained using 4C-CASSCF$^+$ and 4C-CASSCF$^\pm$ orbitals is an order of magnitude smaller than the corresponding difference in the reference energy. To the accuracy of the present calculations, this result indicates that the effects of positive--negative-energy orbital rotations on the CASSCF reference energy and the NVPA PT2 correction are effectively separable.

\subsection{Ground State Correlation in Noble Gas Atoms}
\label{res:1}

In this section, we evaluate the $^1\text{S}_0$ ground-state energies of the all-electron noble-gas atoms from Ne to Xe using 4C-CASCI, 4C-CASSCF$^{+}$, 4C-MRCI and 4C-MRPT2, within NVPA. Calculations were performed with the DC, DCG and DCB Hamiltonians using the Dyall-cv3z basis set for all atoms.\cite{Dyall02_335,Dyall06_441,Dyall12_1217,Dyall16_128,Dyall23_zenodo}

\begin{table*}[ht]
\footnotesize
\centering
\begin{threeparttable}
\centering
\captionsetup{width=\linewidth}
\caption{Correlation energies (in eV) recovered by four-component multireference methods. All CASSCF$^+$ references used an active space of 6 electrons in 8 four-spinors, optimized within the positive-energy manifold. The 4C-MRCI calculations correlated 8 electrons in 34 spinors for Ne and Ar, and 8 electrons in 36 spinors for Kr and Xe. The 4C-MRPT2 calculations correlated all electrons over the full external space. Correlation energies are reported relative to the corresponding 4C-HF energy.}
\label{tab:mrpt1}
\setlength\tabcolsep{12pt}
\begin{tabular}{lllrrrr}
\toprule \toprule
Four-Component & Reference & Correlation & \multirow{2}{*}{Ne} & \multirow{2}{*}{Ar} & \multirow{2}{*}{Kr} & \multirow{2}{*}{Xe} \\
Hamiltonian    &  Method & Method      &                     &                     &                     &                     \\
\midrule
\multirow{5}{*}{DC}    &CASCI & ---   & --0.028  & --0.022  & --0.016  & --0.012 \\
                       &CASCI & MRPT2 & --10.301 & --18.915 & --44.904 & --53.440 \\
                      &CASSCF$^+$  & ---   & --0.253  & --0.082  & --0.057  & --0.026 \\
                       &CASSCF$^+$& MRPT2 & --10.368 & --18.771 & --44.845 & --53.343 \\
                       &CASSCF$^+$ & MRCI  & --4.464  & --4.035  & --3.073  &  --2.439 \\
\midrule
\multirow{5}{*}{DCG}   &CASCI & ---   & --0.028  & --0.022  & --0.016  & --0.012 \\
                       &CASCI & MRPT2 & --10.360 & --19.163 & --46.247 & --56.573 \\
                        & CASSCF$^+$& ---   & --0.253  & --0.082  & --0.057  & --0.026 \\
                       &CASSCF$^+$ & MRPT2 & --10.418 & --19.013 & --46.188 & --56.615 \\
                       &CASSCF$^+$ & MRCI  & --4.468  & --4.041  & --3.071  & --2.445 \\
\midrule
\multirow{5}{*}{DCB}   &CASCI  & ---   & --0.029  & --0.022  & --0.016  & --0.012 \\
                       &CASCI  & MRPT2 & --10.352 & --19.091 & --45.762 & --55.432 \\
                        & CASSCF$^+$&---   & --0.253  & --0.082  & --0.057  & --0.026 \\
                       &CASSCF$^+$ & MRPT2 & --10.417 & --18.944 & --45.701 & --55.473 \\
                       &CASSCF$^+$ & MRCI  & --4.467  & --4.035  & --3.075  & --2.444 \\
\bottomrule \bottomrule
\end{tabular}
\end{threeparttable}
\end{table*}
 
The $^1\text{S}_0$ reference state was described using an active space comprising the occupied $np$ shell and the corresponding unoccupied $(n+1)s$ shell, yielding an active space of six electrons in eight spinors (6e,8o). The reference wavefunctions were generated using either canonical 4C-HF orbitals in 4C-CASCI or state-specific 4C-CASSCF$^+$ orbitals optimized for the $^1\text{S}_0$ ground state. Orbital optimization was restricted to rotations within the positive-energy manifold (4C-CASSCF$^+$). Test calculations in which positive--negative-energy rotations were additionally included (4C-CASSCF$^\pm$) changes the reference and correlation energies by a negligible margin, and are therefore not reported separately. This is consistent with our previous 4C-CASSCF study, in which positive--negative-energy orbital mixing was found to be appreciable only for core and core-excited states. 

Two post-CASSCF correlation treatments are considered for the $^1\text{S}_0$ state: 4C-MRCI and 4C-MRPT2. In both methods, the correlation space is enlarged beyond the complete active space to recover dynamic correlation. The default 4C-MRCI calculations employ the full CI excitation manifold, in which all configurations generated by excitations from the reference wavefunction into the correlation space are included.

 \begin{table}[ht]
\footnotesize
\centering
\begin{threeparttable}
\captionsetup{width=0.45\textwidth}
\caption{Correlation space used in the 4C-MRCI and 4C-MRPT2 calculations for the noble-gas atoms. The number of correlated electrons (Corr. Electrons), correlated positive-energy orbitals (Corr. Orbitals), and resulting determinants (nDets) are listed for each method.}
\label{tab:mrpt2}
\setlength\tabcolsep{3pt}
\begin{tabular}{llccc}
\toprule \toprule
System              & Correlated    & Corr.      & Corr.        & nDets      \\
                    & Method        & Orbitals   & Electrons    &            \\
\midrule
\multirow{2}{*}{Ne} & 4C-MRCI       & 34         &  8   & 1.8 $\times$ 10$^7$         \\
                    & 4C-MRPT2      & 120        &  10  & 2.7 $\times$ 10$^6$       \\

\midrule
\multirow{2}{*}{Ar} & 4C-MRCI       & 34         &  8  & 1.8 $\times$ 10$^7$         \\
                    & 4C-MRPT2      & 194        &  18 & 3.9 $\times$ 10$^7$        \\

\midrule
\multirow{2}{*}{Kr} & 4C-MRCI       & 36         & 8   & 3.0 $\times$ 10$^7$       \\
                    & 4C-MRPT2      & 316        & 36  & 5.4 $\times$ 10$^8$        \\

\midrule
\multirow{2}{*}{Xe} & 4C-MRCI       & 36         & 8   & 3.0 $\times$ 10$^7$       \\
                    & 4C-MRPT2      & 406        & 54  & 2.1 $\times$ 10$^9$       \\
\bottomrule \bottomrule
\end{tabular}
\end{threeparttable}
\end{table}

The correlation energies reported in \Cref{tab:mrpt1} are defined as the energy difference between each correlation calculation and the corresponding 4C-HF energy with the same relativistic Hamiltonian. The corresponding correlation spaces employed in the 4C-MRCI and 4C-MRPT2 calculations are summarized in \Cref{tab:mrpt2}. The 4C-MRCI calculations correlate 8 electrons in 34 positive-energy spinors for Ne and Ar, comprising the active space together with a truncated set of core and low-lying virtual spinors. For Kr and Xe, the correlation space is enlarged to 36 positive-energy spinors to account for the $(n+2)s$ orbitals, which become nearly degenerate with the $(n+1)$ shell orbitals.

For Ne, this represents only about $1/4$ of the complete positive-energy orbital space (120 spinors). In contrast, the low-scaling formulation of 4C-MRPT2 allows the second-order correction to be evaluated over the complete external excitation space generated from all occupied and virtual positive-energy spinors. Consequently, 4C-MRPT2 recovers dynamic correlation up to the second-order from the entire positive-energy orbital space without truncating the virtual or the core manifold.

\Cref{tab:mrpt1} shows that, because of the small (6e,8o) active space, both CASCI and CASSCF$^+$ recover only a small fraction of the total correlation energy. Enlarging the correlation space to (8e,34o) or (8e,36o) in the 4C-MRCI calculations improves the correlation recovery. The recovered correlation energy decreases monotonically from $-4.46$~eV for Ne to $-2.35$~eV for Xe (DC Hamiltonian). This trend indicates that a fixed orbital truncation scheme becomes progressively less effective for heavier elements.

\begin{table*}[ht]
\footnotesize
\centering
\caption{Correlation energies obtained with 4C-MRPT2 and 4C-MRCISD using DC-, DCG-, and DCB-CASSCF$^{+}$ reference wavefunctions. All CASSCF$^{+}$ calculations employed an active space of 6 electrons in 8 spinors (6e,8o) within the NVPA. For Ne, the correlation calculations included 10 electrons in 56 positive-energy spinors. For Ar, Kr, and Xe, 18 electrons in 58 positive-energy spinors were correlated. The 4C-MRCISD and 4C-MRPT2 calculations employed identical correlation spaces and excitation restrictions. All energies are reported in eV.}
\label{tab:pt2vscisd}
\setlength\tabcolsep{10pt}
\begin{tabular}{llrrrr}
\toprule \toprule
Four-Component  & Correlation & \multirow{2}{*}{Ne} & \multirow{2}{*}{Ar} & \multirow{2}{*}{Kr} & \multirow{2}{*}{Xe} \\
Hamiltonian     & Method      &                     &                     &                     &                     \\
\midrule
\multirow{2}{*}{DC}   & MRPT2   & --6.862 & --7.064 & --5.153 & --6.597 \\
                      & MRCISD  & --5.764  & --5.503  & --4.048  &  --5.132 \\
\midrule
\multirow{2}{*}{DCG}  & MRPT2   & --6.862 & --7.082 & --5.160 & --6.451 \\
                      & MRCISD  & --5.770  & --5.510  & --4.054  &  --4.859 \\
\midrule
\multirow{2}{*}{DCB}  & MRPT2   & --6.874 & --7.078 & --5.145 & --6.447 \\
                      & MRCISD  & --5.769  & --5.503  & --4.053  & --4.856 \\
\bottomrule \bottomrule
\end{tabular}
\end{table*}


The 4C-MRPT2 correlation energies are an order of magnitude larger than 4C-MRCI and increase along the series, from $-10.37$~eV ($-0.381$~$E_\text{h}$)
for Ne to $-53.44$~eV ($-1.964$~$E_\text{h}$) for Xe. 
It is worth emphasizing that the missing correlation in 4C-MRCI does not arise from neglecting low-lying virtual spinors, which are already included in the MRCI expansion. Instead, it originates primarily from core--valence pair correlation involving core and high-lying virtual spinors, whose importance increases rapidly with atomic number.

Comparison of the 4C-MRPT2 correlation energies obtained with the DC, DCG, and DCB Hamiltonians shows that the two-electron relativistic contribution beyond Coulomb increases rapidly with atomic number. The difference in correlation energy between the DC and DCB Hamiltonians grows from 0.05~eV for Ne to 2.03~eV for Xe. The Gaunt approximation consistently overestimates the Breit contribution because it neglects the partial cancellation arising from the gauge term. For Xe, this overestimation reaches 1.14~eV.
In contrast, the corresponding DC--DCB differences at the 4C-MRCI level remain below 0.1~eV across the entire series, consistent with our previous 4C-CASSCF study.\cite{Li23_044101} 
The present all-electron 4C-MRPT2 results suggest that, although the Breit contribution to correlation is small within the valence space, it becomes significant for the complete correlation energy, reaching approximately 4\% of the total correlation energy in Xe.
 
The influence of orbital optimization is confined almost entirely to the reference energies (see \Cref{res:0}). At the 4C-MRPT2 level, the dependence of the recovered correlation energy on the choice of reference orbitals is minimal. For the DCB Hamiltonian, the difference in the second-order correlation energy obtained with 4C-HF and 4C-CASSCF$^+$ reference orbitals is consistently below 0.7\%. The extensive external excitation space in 4C-MRPT2 allows the perturbative correction to recover most of the orbital relaxation that is treated variationally in CASSCF. Consequently, the choice of reference orbitals has only a minor effect on the final correlation energy for the systems considered here.

\Cref{tab:pt2vscisd} compares the correlation energies recovered by 4C-MRCISD and 4C-MRPT2 using identical correlation spaces for the DC, DCG, and DCB Hamiltonians. For Ne, the correlation space consists of 10 correlated electrons in 56 positive-energy spinors, including all core orbitals and virtual spinors up to the $4n$ shell. For Ar, Kr, and Xe, the correlation space comprises 18 correlated electrons in 58 positive-energy spinors, including the external $ns$ and $(n-1)d$ core shells together with external virtual spinors spanning the $(n+1)p$ to $(n+2)f$ manifolds. The 4C-MRCISD calculations employ the same excitation restrictions as 4C-MRPT2, permitting at most two holes in the external core space and two electrons in the external virtual space.
Therefore, for atomic systems, 4C-MRCISD provides the variational benchmark for assessing the accuracy of second-order perturbative approximations within the same correlation space.

\Cref{tab:pt2vscisd} shows that 4C-MRPT2, as a non-variational method, overestimates the correlation energy by more than 1~eV for all atoms and all relativistic Hamiltonians considered in this medium-sized correlation space. The magnitude of the overcorrelation increases with atomic number, reaching its largest value for Xe. This behavior reflects the increasingly strong dynamic correlation in heavier atoms, for which a second-order perturbative treatment becomes less accurate. Nevertheless, the overcorrelation remains systematic and nearly independent of the choice of relativistic Hamiltonian.

\subsection{Frozen Core and Virtual Approximations in 4C-MRPT2}
\label{res:2}

In this section we assess the frozen-core and frozen-virtual approximations within 4C-MRPT2, both of which are attractive because the cost of the integral transformation and the perturbative step grows steeply with the number of correlated occupied and virtual spinors. As representative test cases we consider the group-13 atoms gallium and indium, with ground-state configurations [Ar]$3d^{10}4s^2 4p^1$ and [Kr]$5s^2 5p^1$, respectively. 

The reference wavefunctions are generated with 4C-CASSCF$^+$ using an active space of three electrons in eight spinors (3e,8o), corresponding to the valence $ns$ and $np$ shells with the Dyall-v2z basis set.\cite{Dyall98_366,Dyall23_zenodo} Ga and In have 28 and 46 occupied core spinors, respectively.

\begin{figure}[ht]
    \centering
    \includegraphics[width=\linewidth]{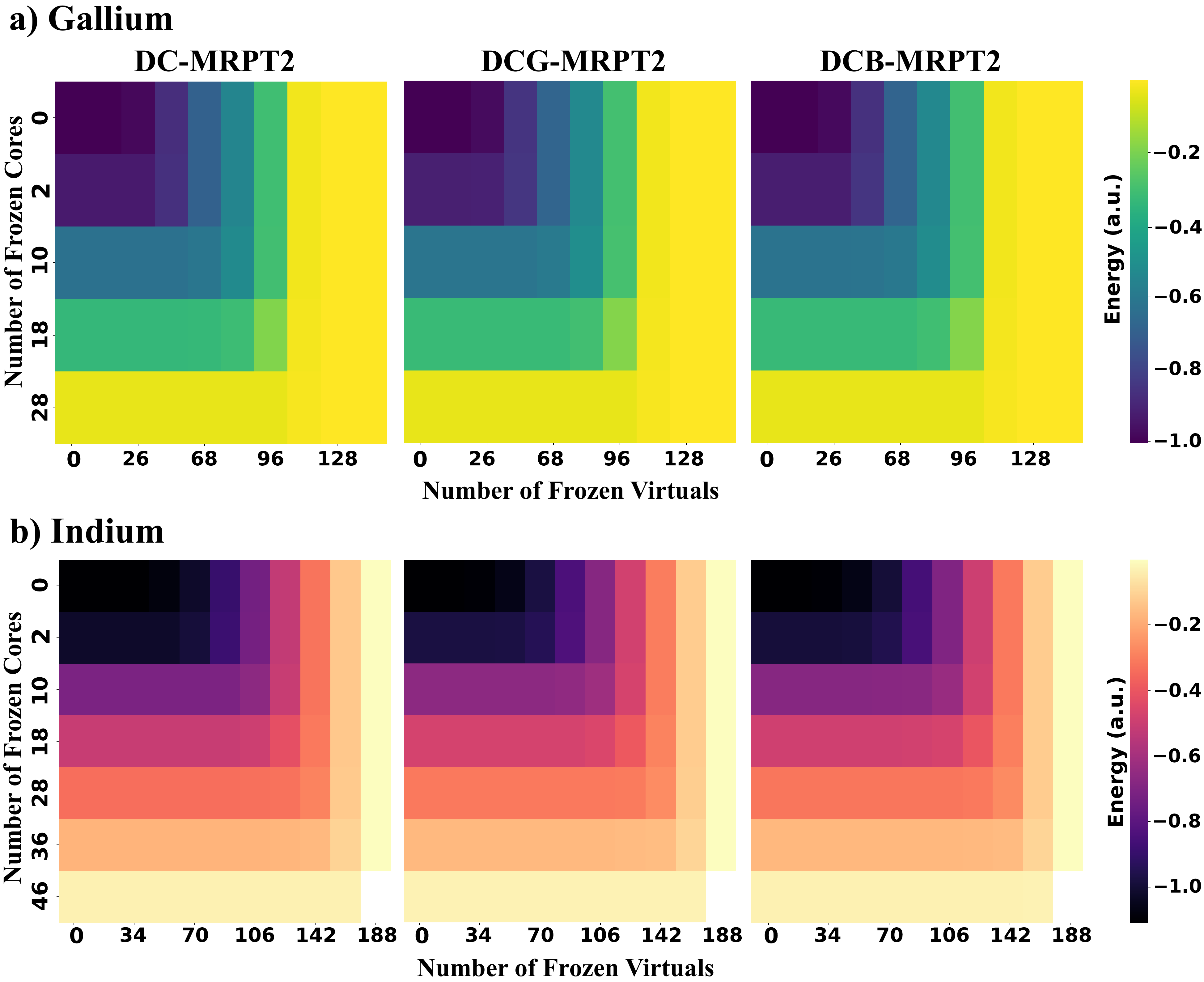}
    \captionsetup{width=\linewidth}
    \caption{Effects of freezing core and virtual molecular spinors using DC-, DCG- and DCB-MRPT2. a) Heatmap demonstrating variations in ground state correlation energies (a.u.) with number of frozen core and virtual spinors for gallium atom. (b) Heatmap demonstrating variations in ground state correlation energies (a.u.) with number of frozen core and virtual spinors for Indium atom.}
    \label{fig:fcv_group13}
\end{figure}

 \begin{figure}[ht]
    \centering
\includegraphics[width=\linewidth]{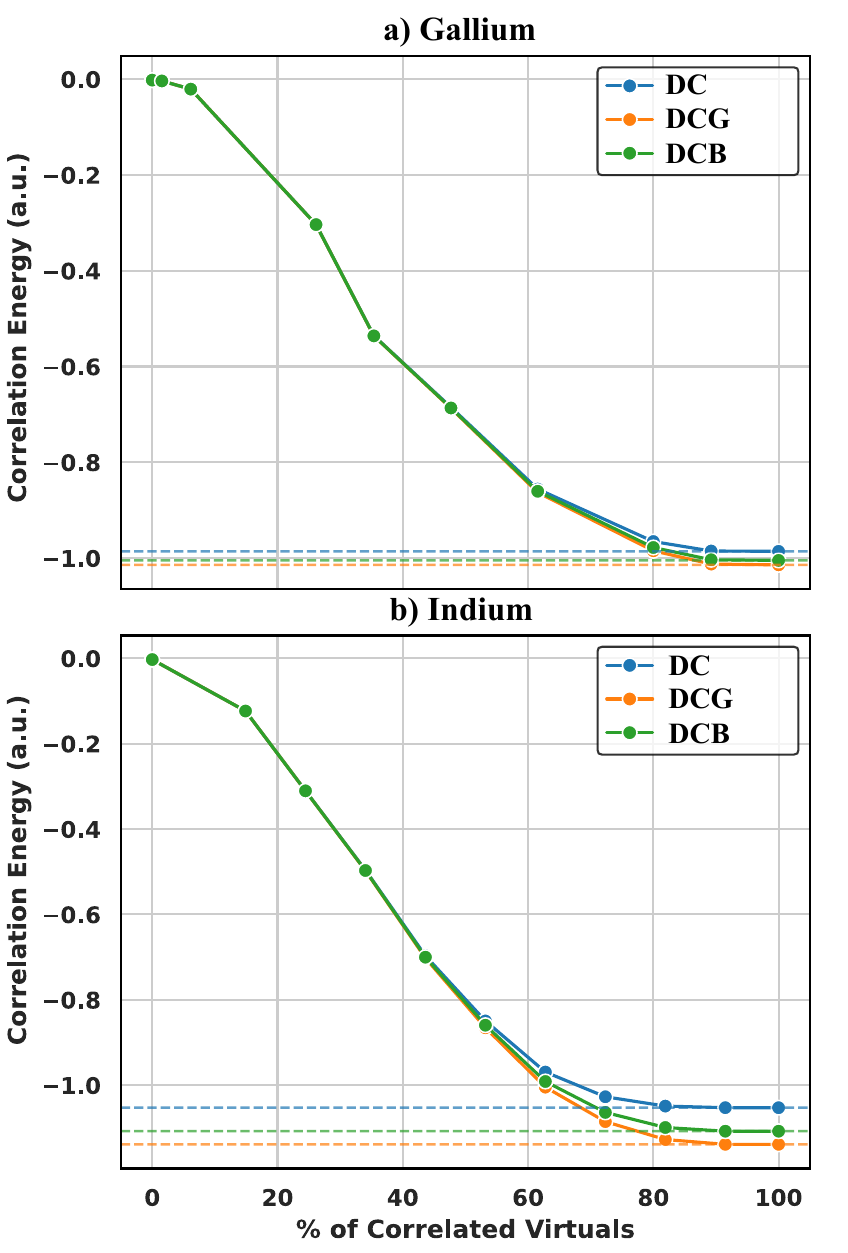}
    \captionsetup{width=\linewidth}
    \caption{Correlation energy (in a.u.) recovered in a) Gallium and b) Indium vs. percentage of virtual spinors included in the external space. The dashed lines denote the untruncated limit for DC-, DCG- and DCB-MRPT2, respectively.}
    \label{fig:fvirt}
\end{figure}

\Cref{fig:fcv_group13} and \Cref{fig:fvirt} present the amount of correlation recovered using frozen spinor approximations in 4C-MRPT2 for the ground states of Ga and In, respectively. In the heat maps the horizontal and vertical axes denote the number of frozen virtual and frozen core spinors, respectively, and the color scale runs from light to dark as the recovered correlation energy approaches the untruncated 4C-MRPT2 limit. The top-left corner therefore corresponds to the full correlation space calculation.
 
For Ga, the recovered correlation energy decreases steadily as core spinors are frozen, with nearly identical behavior for the DC, DCG, and DCB Hamiltonians. Freezing only the $1s$ Kramers pair removes about 7\% of the total second-order correlation energy, reflecting the large but chemically inert core correlation of the deep $1s$ shell. As additional core shells are frozen, the recovered correlation energy continues to decrease. When the entire [Ar]$3d^{10}$ core is excluded, only about 35\% of the total correlation energy remains. The $3d^{10}$ shell alone contributes roughly 33\%, highlighting the importance of $(n-1)d$--$np$ core--valence correlation in heavier group-13 elements.

The frozen-virtual approximation is considerably more forgiving. Excluding the 20\% of the highest-energy virtual spinors changes the 4C-MRPT2 correlation energy by less than 1\%, indicating that these high-energy spinors contribute little to the first-order wavefunction. Beyond a 40\% truncation, however, the recovered correlation energy deteriorates rapidly (\Cref{fig:fvirt}). Since the number of discarded virtual spinors depends on the basis set, the corresponding orbital-energy cutoff provides a more transferable criterion: the 20\% threshold corresponds to approximately 0.020~$E_{\mathrm{h}}$ for both Ga and In.

\begin{table}[ht]
\footnotesize
\centering
\begin{threeparttable}
\captionsetup{justification=justified}
\caption{Fine-structure splittings in ground electronic state (${}^2\text{P}$) of Gallium and Indium. Values in parentheses are percentage deviations from experiment. The experimental data from latest literature are also shown. The units are in cm$^{-1}$.}
\label{tab:mrpt3}
\setlength\tabcolsep{6pt}
\begin{tabular}{llcc}
\toprule \toprule
4C-       & Correlation & \multirow{2}{*}{Ga} & \multirow{2}{*}{In} \\
Method    & Method      &                     &          \\
\midrule
\multirow{2}{*}{DC} & CASSCF$^+$       & 724 (12.4) &  1905 (13.9)  \\
                    & MRPT2            & 838 (1.4) & 2220 (0.3) \\
\midrule
\multirow{2}{*}{DCG} & CASSCF$^+$       & 709 (14.1) &  1878 (15.1) \\
                     & MRPT2            & 820 (0.7) & 2118 (4.3) \\
\midrule
\multirow{2}{*}{DCB} & CASSCF$^+$       & 711 (13.9) &  1917 (13.3) \\
                     &  MRPT2           & 820 (0.7) & 2146 (3.0) \\
\midrule
\multicolumn{2}{l}{Exp\tnote{a}} & 826 & 2213 \\
\bottomrule \bottomrule
\end{tabular}
\begin{tablenotes}[flushleft]
\footnotesize
\item[a] Ga, Ref.~\citenum{Sugar07_509}; In, Ref.~\citenum{Litzen01_4475}.
\end{tablenotes}
\end{threeparttable}
\end{table}
The spin--orbit splitting of the ${}^2\mathrm{P}$ ground state in the group-13 atoms provides a stringent test of the interplay between static and dynamic correlation in fully relativistic calculations of chemically relevant properties. Spin--orbit coupling splits this state into the ${}^2\text{P}_{1/2}$ and ${}^2\text{P}_{3/2}$ fine-structure levels, separated by 826 cm$^{-1}$ in Ga\cite{Sugar07_509} and 2213 cm$^{-1}$ in In, measured experimentally.\cite{Litzen01_4475} The corresponding 4C-CASSCF$^+$ and 4C-MRPT2 splittings, obtained with an active space of (3e,8o) (the $ns$ and $np$ valence shells) and the DC, DCG, and DCB Hamiltonians, are collected in \cref{tab:mrpt3}. The 4C-CASSCF$^+$ reference wavefunctions for the ${}^2\mathrm{P}{1/2}$ and ${}^2\mathrm{P}{3/2}$ states were optimized using a state-averaged CASSCF procedure. The subsequent PT2 calculations were performed separately for each state using the corresponding state-specific reference wavefunction.
 
At the CASSCF$^+$ level, the fine-structure splitting is underestimated by 12--15\% for both Ga and In. This error is nearly independent of the relativistic two-electron Hamiltonian: the calculated splittings are 724, 709, and 711~cm$^{-1}$ for Ga, and 1905, 1878, and 1917~cm$^{-1}$ for In, using the DC, DCG, and DCB Hamiltonians, respectively. The nearly identical splittings indicate that the error does not arise from the treatment of the relativistic two-electron interaction. Instead, it reflects the omission of dynamic correlation outside the valence active space, particularly from the $(n-1)d$ and outer-core shells.
 
Including dynamic correlation through 4C-MRPT2 improves the fine-structure splitting by approximately 110~cm$^{-1}$ for Ga and 230--315~cm$^{-1}$ for In, bringing the calculated values into excellent agreement with experiment. For In, however, the near-perfect agreement obtained with the DC Hamiltonian should be interpreted with caution. It arises from a fortuitous cancellation between the neglected two-electron spin--orbit screening and the remaining errors due to incomplete correlation recovery and basis-set incompleteness.
 
Taken together, these results establish that while the 4C-CASSCF$^+$ reference recovers the qualitative fine-structure pattern, external-space correlation introduced by 4C-MRPT2 is essential for quantitative accuracy, bringing the splittings of both atoms to within a few percent of the experiment with the most complete DCB Hamiltonian.

\section{Conclusion}
\label{conclusion}
We have developed a four-component multireference second-order perturbation theory (4C-MRPT2) within the STP-DAS framework. The method combines a variational four-component multiconfigurational reference with a second-order perturbative treatment of dynamic correlation, while remaining fully compatible with the Dirac--Coulomb (DC), Dirac--Coulomb--Gaunt (DCG), and Dirac--Coulomb--Breit (DCB) Hamiltonians. By extending the STP-DAS formalism to the perturbative correction, the method retains the memory efficiency and massive parallel scalability previously demonstrated for large-scale configuration interaction calculations. 

Benchmark calculations reveal several important physical insights. The influence of positive--negative-energy orbital rotations is largely confined to the multiconfigurational reference, while the second-order correlation energy is only weakly dependent on the choice of reference orbitals. We also show that the Breit contribution to the all-electron correlation energy increases rapidly with atomic number and is substantially underestimated when the correlation treatment is restricted to truncated orbital spaces. Furthermore, the Gaunt approximation systematically overestimates the Breit contribution because it neglects the partial cancellation arising from the gauge term.

The present implementation also provides practical strategies for reducing computational cost. Frozen-core and frozen-virtual approximations introduce only modest errors while significantly reducing the size of the perturbative space, making all-electron four-component correlation calculations substantially more affordable. At the same time, the computational bottleneck shifts increasingly toward the relativistic AO-to-MO integral transformation as more complete two-electron Hamiltonians are employed.

4C-MRPT2 significantly extends the range of molecular systems accessible to accurate four-component electronic structure theory while providing a practical platform for future developments, including excited-state methods, spectroscopic properties, and large-scale applications involving heavy-element molecules and materials.

\begin{acknowledgement}
The development of variational relativistic multi-reference methods is supported by the U.S. Department of Energy, Office of Science, Basic Energy Sciences, in the Computational and Theoretical Chemistry program (Grant No. DE-SC0006863 to XL). The development of the Chronus Quantum computational software is supported by the Office of Advanced Cyberinfrastructure, National Science Foundation (Grants No. OAC-2103717).
This work used the Hyak supercomputer at the University of Washington and the Bridges-2 system at the Pittsburgh Supercomputing Center through the NSF ACCESS program.
\end{acknowledgement}

\newpage
\bibliography{
    References/Journal_Short_Name,
    References/relativity_intro,
    References/mrpt,
    References/ci,
    References/Li_Group_References,
    References/misc,
    References/basisset,
    References/section3_results
}


%
%

\end{document}